\documentclass[12pt,preprint]{aastex}
\usepackage[modulo, switch]{lineno}

\newcommand{\dsct}{$\delta$~Scuti }

\newcommand{\Msun}{$M_{\odot}$}
\newcommand{\Lsun}{$L_{\odot}$}

\newcommand{\vsini}{$v\cdot\sin{i}$}
\newcommand{\cd}{d$^{\rm -1}$}

\shorttitle{Pulsational analysis of V 588 Mon and V 589 Mon}
\shortauthors{Zwintz et al.}

\begin{document}

\title{Pulsational analysis of V 588 Mon and V 589 Mon observed with the MOST\thanks{Based on data from the {\it MOST} satellite, a Canadian Space Agency mission, jointly operated by Dynacon Inc., the University of Toronto Institute for Aerospace Studies and the University of British Columbia with the assistance of the University of Vienna.}  and CoRoT\thanks{The CoRoT space mission, launched on December 27th 2006, has been developed and is operated by CNES, with the contribution of Austria, Belgium, Brazil , ESA (RSSD and Science Programme), Germany and Spain.} satellites}


\author{K. Zwintz}
\affil{Institute of Astronomy, T\"urkenschanzstrasse 17, A-1180 Vienna, Austria}
\email{konstanze.zwintz@univie.ac.at}

\author{T. Kallinger}
\affil{Department of Physics and Astronomy, University of British Columbia, 6224 Agricultural Road, Vancouver, BC V6T 1Z1, Canada}

\author{D. B. Guenther and M. Gruberbauer}
\affil{Department of Astronomy and Physics, St. Mary's University, Halifax, NS B3H 3C3, Canada}

\author{R. Kuschnig and W. W. Weiss}
\affil{Institute of Astronomy, T\"urkenschanzstrasse 17, A-1180 Vienna, Austria}

\author{M. Auvergne}
\affil{LESIA, Observatoire de Paris-Meudon, 5 place Jules Janssen, 92195 Meudon, France}

\author{L. Jorda}
\affil{Laboratoire d'Astrophysique de Marseille, P\^{o}le de l'\'{E}toile Site de Ch\^{a}teau-Gombert, 38, rue Fr\'{e}d\'{e}ric Joliot-Curie, 13388 Marseille, France}

\author{F. Favata}
\affil{European Space Agency, 8-10 rue Mario Nikis, 75015 Paris, France}

\author{J. Matthews}
\affil{Department of Physics and Astronomy, University of British Columbia, 6224 Agricultural Road, Vancouver, BC V6T 1Z1, Canada}

\and

\author{M. Fischer}
\affil{Vienna University of Technology, Institute of Communications and Radio-Frequency Engineering, Gusshausstrasse 25/389, A-1040 Vienna, Austria}

\begin{abstract}
The two pulsating pre-main sequence (PMS) stars V 588 Mon and V 589 Mon were observed by CoRoT for 23.4 days in March 2008 during the Short Run SRa01 and in 2004 and 2006 by MOST for a total of $\sim$70 days. We present their photometric variability up to 1000 $\mu$ Hz and down to residual amplitude noise levels of 23 and 10 ppm of the CoRoT data for V 588 Mon and V 589 Mon, respectively. The CoRoT imagette data as well as the two MOST data sets allowed for detailed frequency analyses using Period04 and SigSpec. 
We confirm all previously identified frequencies, improve the known pulsation spectra to a total of 21 frequencies for V 588 Mon and 37 for V 589 Mon and compare them to our PMS model predictions. No model oscillation spectrum with $l$ = 0, 1, 2, and 3 $p$-modes matches all the observed frequencies. 
When rotation is included we find that the rotationally split modes of the slower rotating star, V 589 Mon, are addressable via perturbative methods while for the more rapidly rotating star, V 588 Mon, they are not and, consequently, will require more sophisticated modeling. The high precision of the CoRoT data allowed us to investigate the large density of frequencies found in the region from 0 to 300 $\mu$Hz. The presence of granulation appears to be a more attractive explanation than the excitation of high-degree modes.
Granulation was modeled with a superposition of white noise, a sum of Lorentzian-like functions and a Gaussian. 
Our analysis clearly illustrates the need for a more sophisticated granulation model.
\end{abstract}





\section{Introduction}
V 588 Mon (HD 261331, NGC 2264 2) and V 589 Mon (HD 261446, NGC 2264 20) are two pre-main-sequence (PMS) pulsating stars for which there exists strong evidence that they are members of the young open cluster NGC 2264. The proper motions for both stars are in agreement with the clusterÕs average proper motion \citep{hog00} and both fit the clusterÕs HR- and color-magnitude diagrams well. A radial velocity measurement only exists for V 589 Mon \citep{str71} but it is consistent with the values for other cluster members. 
Finally, the radial velocities of emission lines in the optical spectra of the two stars caused by interstellar gas match the radial velocities of the cluster suggesting that both stars are indeed embedded in gas clouds belonging to the cluster \citep{kal08}.


The cluster has a diameter of $\sim$39 arcminutes and belongs to the Mon OB 1 association. The age of NGC 2264 is reported in the literature to lie between 3 \citep[e.g.,][]{wal56,sun04} and 10 million years \citep[e.g.,][]{sag86}. With such a young age, the cluster's main sequence only consists of massive O and B type stars, while stars of later spectral types are still in their pre-main sequence (PMS) phase. Therefore, V 588 Mon and V 589 Mon having spectral types of A7 and F2, respectively, have not arrived on the zero-age main sequence (ZAMS) yet. 

The \dsct like variability of V 588 Mon and V 589 Mon was first reported by \citet{bre72}. Hence, they were the first pulsating PMS stars discovered. In the meantime the number of known \dsct like PMS pulsators has increased from 36 \citep{zwi08} to $\sim$ 60 (Zwintz, K., private communication) due to dedicated observations from ground and from space.

Pulsating PMS stars have intermediate masses, i.e. between $\sim$1.5 and 4 M$_{\odot}$ and can become vibrationally unstable when they cross the instability region in the Hertzsprung-Russell (HR) diagram on their way to the ZAMS. Pre- and post-main sequence evolutionary tracks for the same stellar mass intersect several times close to the ZAMS which makes the determination of the evolutionary stage of a field star from only its effective temperature, luminosity and mass ambiguous. Additional information such as typical observational evidence for the PMS evolutionary stage (i.e. emission lines, IR excess, an X-ray flux, located in an obscured region on the sky etc.) or membership to very young open clusters is needed to resolve this ambiguity. Another way to distinguish the evolutionary stages can come from the asteroseismic interpretation of the observed pulsation frequencies \citep{gue07}.

Time series photometry for V 588 Mon and V 589 Mon has been obtained from a multi-site ground based campaign \citep{kal08} and from two observing runs of the Canadian MOST space telescope \citep{wal03} in 2004 and 2006 \citep{gue09}. The 8 and 12 frequencies common to these three data sets for V 588 Mon and V 589 Mon, respectively, were submitted to a first asteroseismic analysis \citep{gue09}. The accuracy of the MOST observations is higher than that of the ground based data. Hence it is not surprising that the two MOST data sets yield more significant frequencies at lower amplitudes that were not found in the ground based observations.
The CoRoT observations of the two PMS pulsators are new and independent data sets of unprecedented accuracy that allow us for the first time to investigate other effects (e.g., granulation) in PMS stars. As both MOST data sets and the CoRoT data are available to us, we use them together for a detailed pulsational analysis.

\section{Observations and Data Reduction}

\subsection{MOST observations}

The MOST (Microvariability and Oscillations of STars) space telescope \citep{wal03} was launched on June 30, 2003, into a polar sun-synchronous circular orbit with an orbital period of $\sim 101$ minutes (corresponding to an orbital frequency of 14.2 \cd). From its orbital vantage point, MOST can obtain uninterrupted observations of stars located in its Continuous Viewing Zone (CVZ) for up to 8 weeks. The MOST satellite houses a 15-cm Rumak-Maksutov telescope feeding a CCD photometer through a single, custom broadband optical filter (covering wavelengths from 350 to 750\,nm).

MOST can supply up to three types of photometric data simultaneously for multiple targets in its field.  The mission was originally intended only for Fabry Imaging, in which an image of the entrance pupil of the telescope - illuminated by a bright target star ($V < 6$) - is projected onto the instrument's Science CCD by a Fabry microlens \citep[see][]{ree06} for details. After MOST was operating in orbit, the pointing performance of the satellite was improved so much that a new mode of observing, Direct Imaging, was made practical.  Direct Imaging is much like conventional CCD photometry, in which photometry is obtained from defocussed images of stars in the open area of the CCD not covered by the Fabry microlens array field stop mask.  In the original design, no scientific information was available from the Guide Stars used for the ACS (Attitude Control System), but now precise photometry is possible for these stars as well \citep[see e.g., ][]{wal05,aer06}.

\begin{figure}[htb]
\centering
\includegraphics[width=0.9\textwidth]{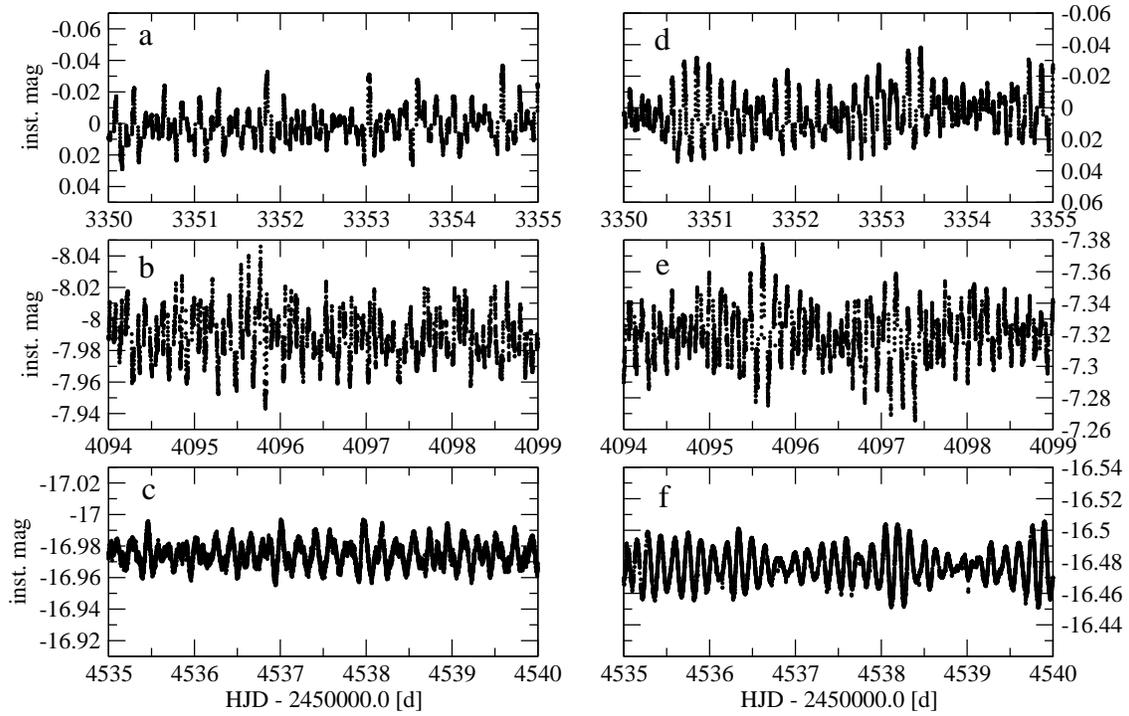}
\caption{5-day subsets of the MOST 2004 (a,d), MOST 2006 (b,e) and CoRoT (c,f) light curves for V 588 Mon (left) and V 589 Mon (right) to the same y-axis scale. The different amplitudes in the MOST versus the CoRoT light curves is mainly due to the different bandpasses used in the two satellites.}
\label{lcs}
\end{figure}

V 588 Mon and V 589 Mon were observed in the Direct Imaging mode, first from December 6, 2004, to January 24, 2005, and second within the NGC 2264 observations \citep[see][]{zwi09} from December 7, 2006, to January 4, 2007. The light curves have therefore time bases of 48.13 days in 2004 and 22.77 days in 2006. In 2004 on-board exposures were 15 seconds long sampled 2 times per minute. A 5-day subset light curve of V 588 Mon from 2004 is shown in panel a of Figure \ref{lcs}, the respective V 589 Mon 5-day subset light curve in panel d.
The 2006 data had exposure times of 1.5 seconds, 16 consecutive images were added on top of each other and the sampling time was 24 seconds \citep{zwi09}. 5-day subsets of the light curves for V 588 Mon and V 589 Mon from the 2006 data are given in panels b and e of Figure \ref{lcs}, respectively.

Data reduction of the MOST Direct Imaging photometry was conducted using the method developed by \citet{row06} which combines classical aperture photometry and point-spread function fitting to the Direct Imaging subrasters.

\subsection{CoRoT observations}
The CoRoT satellite \citep{bag06} was launched on December 27, 2006, from Baikonur aboard a Soyuz rocket into a polar, inertial circular orbit at an altitude of 896 km. With its 27-cm telescope, CoRoT can observe stars within a field of view of about $1.3 \times 2.6 \deg^2$ located inside two cones of 10-degree radius, one at a right ascension of 06:50 (galactic anticenter direction) and the other at 18:50 (galactic center direction).

The CoRoT space telescope originally had two CCDs devoted to asteroseismology for stars with $5.7 < m_V < 9.5$ mag and two CCDs dedicated to the search for exoplanets where $\sim$6000 stars in the magnitude range from 10 to 16 mag in $R$ per CCD are monitored. 
Observations of the young open cluster NGC 2264 were conducted as dedicated Short Run, SRa01, uninterrupted for 23.4 days in March 2008 within the framework of the {\it Additional Programme} \citep[{\it AP;}][]{wei06}. At the middle of SRa01 the ascending mode of the orbital plane was 15.67$^{\circ}$.
The complete cluster was placed in one of the two Exofield-CCDs and data were taken for all stars in the accessible magnitude range. The 100 observed brightest stars in the field of the cluster were primary targets to search for stellar pulsations in PMS stars. A detailed description will be given in Zwintz et al. (2010, in preparation).

With $V$=9.7 mag and $V$=10.3 mag, V 588 Mon and V 589 Mon are nominally too bright to be observed in the CoRoT Exofield. Obtaining data for these two stars was only possible using the method of the so-called CoRoT imagettes.

Per observing run up to 20 imagettes, $15 \times 10$ pixel large CCD subwindows, can be defined in each of the two CoRoT Exo-field CCDs. 
SRa01 was the first CoRoT observing run, for which the imagette data were requested for scientific use and allowed to observe stars that normally would be too bright for the Exo-field. 
Each of the imagettes is submitted to a special data reduction pipeline developed by the Laboratoire d'Astronomie de Marseille (LAM) and the Laboratoire d' \'{E}tudes Spatiales et d'Instrumentation en Astrophysique (LESIA). The reduction includes the flagging of data obtained during passes of the satellite through the South Atlantic Anomaly (SAA), the calculation of a photometric mask, the detection of cosmics, contaminants, outliers due to the satellite jitter and hot pixels and filtering of the orbital signal. A detailed description will be given in Auvergne et al. (2010, in preparation).

In preparation for the NGC 2264 observations, a prioritized target catalog consisting only of the stars brighter than the Exo-CCD magnitude limit was generated. From the 20 possible windows on the Exo-field CCD1, 14 were used to observe some of the previously selected objects including the two known PMS pulsators V 588 Mon and V 589 Mon. Figure \ref{lcs} shows 5-day subsets of the respective light curves (panels c and f). Note that the different filter bandwidths of the CoRoT (from 370 to 1000 nm) compared to the MOST (from 350 to 750 nm) data are caused by the different passbands used in the two satellites.

\subsection{Frequency Analysis}
Frequency analyses were conducted using Period04 \citep{len05} which combines Fourier and least-squares algorithms. 
Frequencies are prewhitened subsequently and are considered to be significant if their amplitudes exceed 4 times the local noise level in the amplitude spectrum \citep{bre93,kus97}. 

For the MOST data sets, the formally significant frequencies were checked against the instrumental frequencies related to the orbit of the satellite, its harmonics and 1 \cd\,\,sidelobes within the frequency resolution \citep[computed according to][]{kall08}. In the CoRoT data the influence of the satellite's orbital and related frequencies is negligible.

\section{Pulsation and granulation of V 588 Mon and V 589 Mon}

\subsection{Modeling the stellar background in the CoRoT data}

Assuming white, i.e. frequency independent, background noise, the 23.4 days long CoRoT time series for V 588 Mon and V 589 Mon showed 106 and 197 formally significant frequencies, respectively, after the initial frequency analysis with the methods described above. Although the number of frequencies are lower than the many hundreds of frequencies reported for classical Scuti stars observed by CoRoT \citep[e.g.,][observed $>$1000 frequencies in HD 50844]{por09}, they are still too many to be accounted for as simply individually excited low $l$-valued $p$-modes.  Asteroseismic models predict some tens of $p$-modes with degrees $l < 4$, which are potentially excited in the frequency range of $\delta$ Scuti-type oscillations.  A substantially larger number of modes would have to include modes of degree $l > 4$ and/or rotationally split modes and/or the possibility that the detected modes are part of the intrinsic background noise due to granulation. 

If all $l$-valued $p$-modes are driven to similar amplitudes then the geometric cancelation effect reduces the observable amplitude by more than an order of magnitude for $l > 3$ $p$-modes \citep[see, for example, the spatial response functions in][]{chr82}. But \citet{das06} note, though, that some stars exhibit more than two orders of magnitude variations in pulsation amplitudes. In addition, when observing in luminosity the geometric cancelation effect is offset by an $l^2$ factor. Regardless, the variation in amplitude of the modes in the short run data (23 d) from CoRoT and in the MOST data is a little more than one order of magnitude. We are not yet reaching the lower amplitude modes, for which these effects become important. Furthermore, we note that unlike the example of HD 50844 observed by \citet{por09} we do have data for V 588 Mon and V 589 Mon from different instruments obtained at different epochs. If all the frequencies were due to pulsation modes then they (or at least the majority) should appear in all data sets. 
However, apart from the frequencies listed in Tables \ref{v588-fs} and \ref{v589-fs} we find 5 to 10 frequencies that coincide (within the frequency uncertainties) in the CoRoT and MOST data sets for both stars. The number depends on the considered frequency range, the chosen significance limits, and how the coincidence and  frequency error is defined in detail. But even for two totally unrelated data sets it would not be surprising to find a few coincident frequencies. In case of V 588 Mon, for example, most of the 106 formally significant frequencies in the CoRoT data are distributed between about 5 and 15 \cd\,\, in about 700 independent frequency bins. The frequency bin-size is one over the data set length times the square root of the significance \citep{kall08}.  For our analysis we use an average significance of 10. Even if the formally significant frequencies in the CoRoT data set were just random numbers one could expect to find by chance matches for at least a few of them with the 179 formally significant frequencies determined from the MOST data. Hence, at this time, we argue that many of the low amplitude frequencies in the CoRoT data are due to granulation and a (rare) coincidence with low amplitude MOST frequencies is not indicative for pulsation.

\citet{kal10} concluded that many of the low-amplitude frequencies in $\delta$ Scuti stars are consistent with a strongly frequency-dependent intrinsic background signal due to granulation or some effect similar to granulation. 
This is well known for cool stars with convective envelopes. The turbulent motions in their outer convective envelopes generate quasi-stochastic power, with amplitudes strongly decreasing for shorter time scales. Although the signal is quasi-stochastic in the time domain, its Fourier transform is Lorentzian-like, where different physical processes on (or near) the stellar surface produce the same type of signal but on different amplitude and time scales. For the Sun and other Sun-like stars, the different signal components are usually assigned to stellar activity, activity of the photospheric/chromospheric magnetic network, and granulation, with time scales ranging from months for active regions to minutes for granulation \citep[see e.g.,][]{mic08,mic09}.

The presence of a granulation background signal depends of course on the presence of a surface convection zone, which is usually attributed to stars cooler than the red border of the classical instability strip. On the other hand, it is known that stellar evolutionary models for stars in the instability strip predict a thin convective surface layer and they should therefore show a similar signature of granulation as cool stars in their power spectra. We believe that this was not recognized before in classical $\delta$ Scuti stars, but also in PMS $\delta$ Scuti stars because the amplitudes are small and a clear detection requires long and uninterrupted observations to achieve a high-frequency instrumental noise well below 100ppm, which have become available only recently. Although the MOST observations of V 588 and V 589 Mon are of unprecedented length and completeness the granulation signal with a low-frequency amplitude of roughly 100ppm (see Figure \ref{v588gran} and \ref{v589gran}) is hidden in the instrumental noise (see Table \ref{stat}). Only the CoRoT observations are sufficient to reveal the strongly frequency-dependent background signal.

\subsubsection{V 588 Mon}

The CoRoT light curve of V 588 Mon consists of 56978 data points sampled with 32 seconds and has a noise level of 47 ppm in the original amplitude spectrum from 0 to 100 \cd (see Table \ref{stat}). The first frequency analysis resulted in 106 formally significant frequencies with signal-to-noise values larger than 4. Among those are also the previously published 8 pulsation frequencies \citep{gue09}.

\begin{figure}[htb]
\centering
\includegraphics[width=0.7\textwidth]{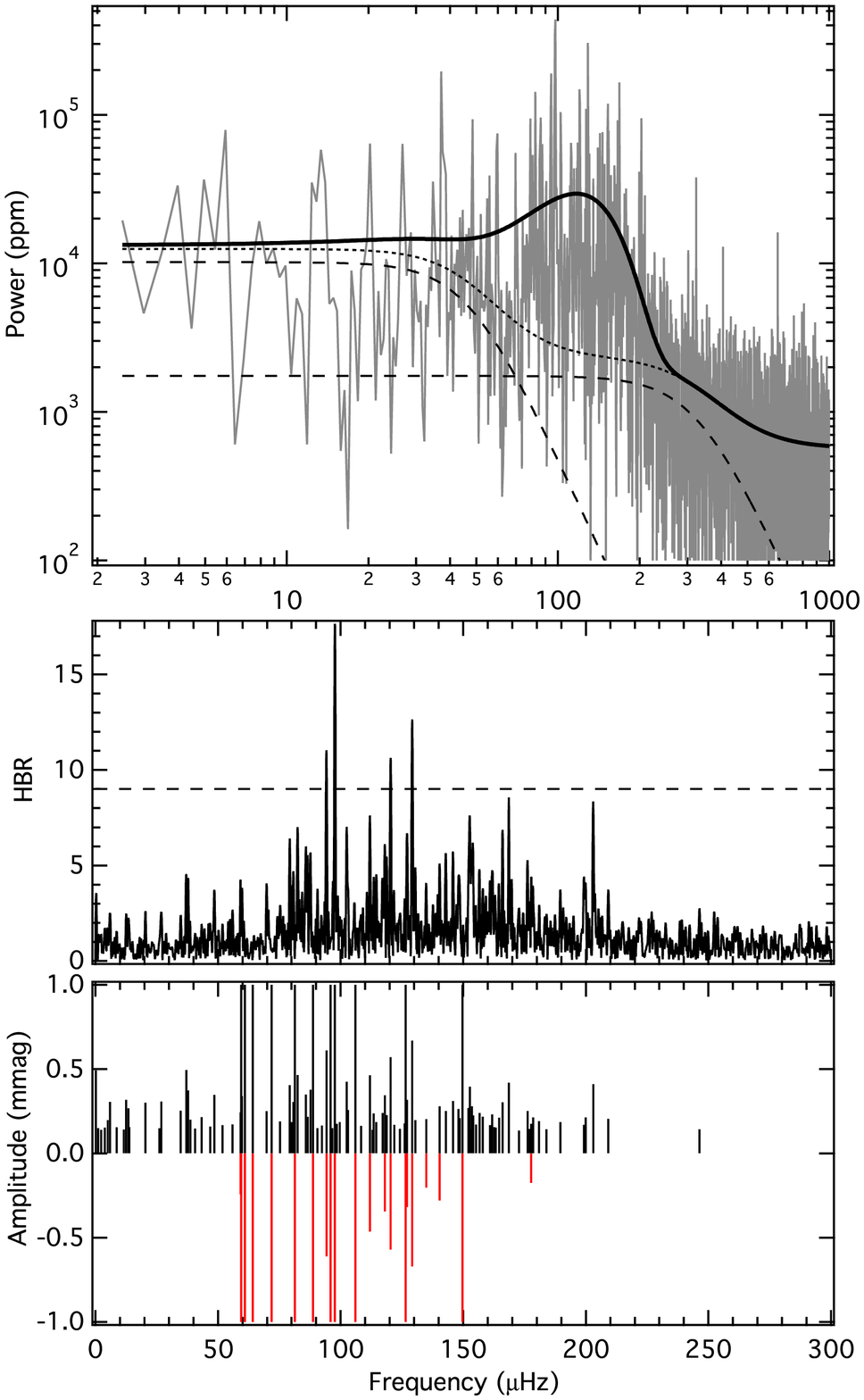}
\caption{{\it Top panel}: Residual gap-filled power spectrum of V 588 Mon (grey) after prewhitening the 10 highest amplitude frequencies. The global fit is shown as solid line, while the dotted line illustrates the model without the pulsational components consisting of the two Lorentzian-like functions and white noise, and the dashed lines mark the two Lorentzian-like functions. {\it Middle panel}: Residual power spectra normalised to the background components, where the dashed line marks a height-to-background ratio (HBR) of 9 (given in power). {\it Bottom panel}: Comparison between the 106 formally significant frequencies (positive values) to the identified 21 pulsation frequencies (negative values) which consist of the 10 prewhitened peaks before granulation modeling, 4 peaks exceeding an HBR value of 9 and 7 peaks common to all available data sets.}
\label{v588gran}
\end{figure}

A visual inspection of the power spectrum of V 588 Mon reveals that the average power per frequency bin (e.g. 10\,$\mu$Hz) is roughly constant at low frequencies and drops by more than one magnitude beyond about 500\,$\mu$Hz, which is a clear indication for a frequency-dependent noise. To model the background signal, the pulsation power has to be taken into account as well. This is relatively straight forward for cool stars, where the envelope of the pulsation power excess is well approximated by a Gaussian. This is more complicated for stars in the instability strip where the excess envelope can be quite different from a Gaussian (or any other simple function). We follow the approach of \citet{kal10} and first prewhiten the 10 frequencies with the most significant peaks (i.e., the highest amplitudes) and assume the residual pulsation power excess to be at least Gaussian like. Although NGC 2264 was observed continuously for the 23.4 days, several data points were removed during the reduction process mostly due to high energy particle hits during the satelliteÕs passes through the South Atlantic Anomaly (SAA) leading to regular gaps in the time series. 
The resulting alias peaks would significantly distort a fit to the power spectrum. We therefore filled the gaps -- but only for the time series used to investigate the background characteristics which requires uninterrupted data -- with linearly interpolated values to receive the needed clean window function. 
The power spectrum of the gap-filled data is shown in Figure \ref{v588gran}. The residual power spectrum is modeled with a superposition of white noise, two Lorentzian-like functions, and a Gaussian. The resulting global fit (solid line in Figure \ref{v588gran}) reproduces the overall shape of the residual power spectrum and demonstrates the strong frequency dependence of the background signal. The residual power spectrum is then corrected for the background signal by using the global fit without the pulsation component. Finally, we use the height-to-background ratio (HBR; middle panel of Figure \ref{v588gran}) to rate the significance of the individual frequencies received from the initial frequency analysis and consider only frequencies that exceed an HBR value of 9 to originate from pulsation. For V 588 Mon four frequencies match this criterion (see Table  \ref{v588-fs} and Figure \ref{v588gran}). Together with the 10 prewhitened frequencies of highest amplitudes, a total of 14 pulsation frequencies remain after granulation modeling. There are a number of peaks in the middle panel of Figure \ref{v588gran} between about 70 and 200\,$\mu$Hz that have a higher HBR value than the peaks in the surrounding frequency ranges but which do not exceed our formal limit of HBR $\geq$ 9. This indicates that there are additional pulsation frequencies but our simplistic model of the pulsation power excess underestimates their significance.

\subsubsection{V 589 Mon}

The CoRoT light curve of V 589 Mon has 57092 data points also obtained with an exposure time of 32 seconds and the noise level in the amplitude spectrum of the original data set is 61 ppm from 0 to 100 \cd (Table \ref{stat}). After the initial frequency analysis, 197 frequencies were identified to be formally significant. Similar as for V 588 Mon, all previously published 12 pulsation frequencies \citep{gue09} were also detected in the CoRoT data.

\begin{figure}[htb]
\centering
\includegraphics[width=0.7\textwidth]{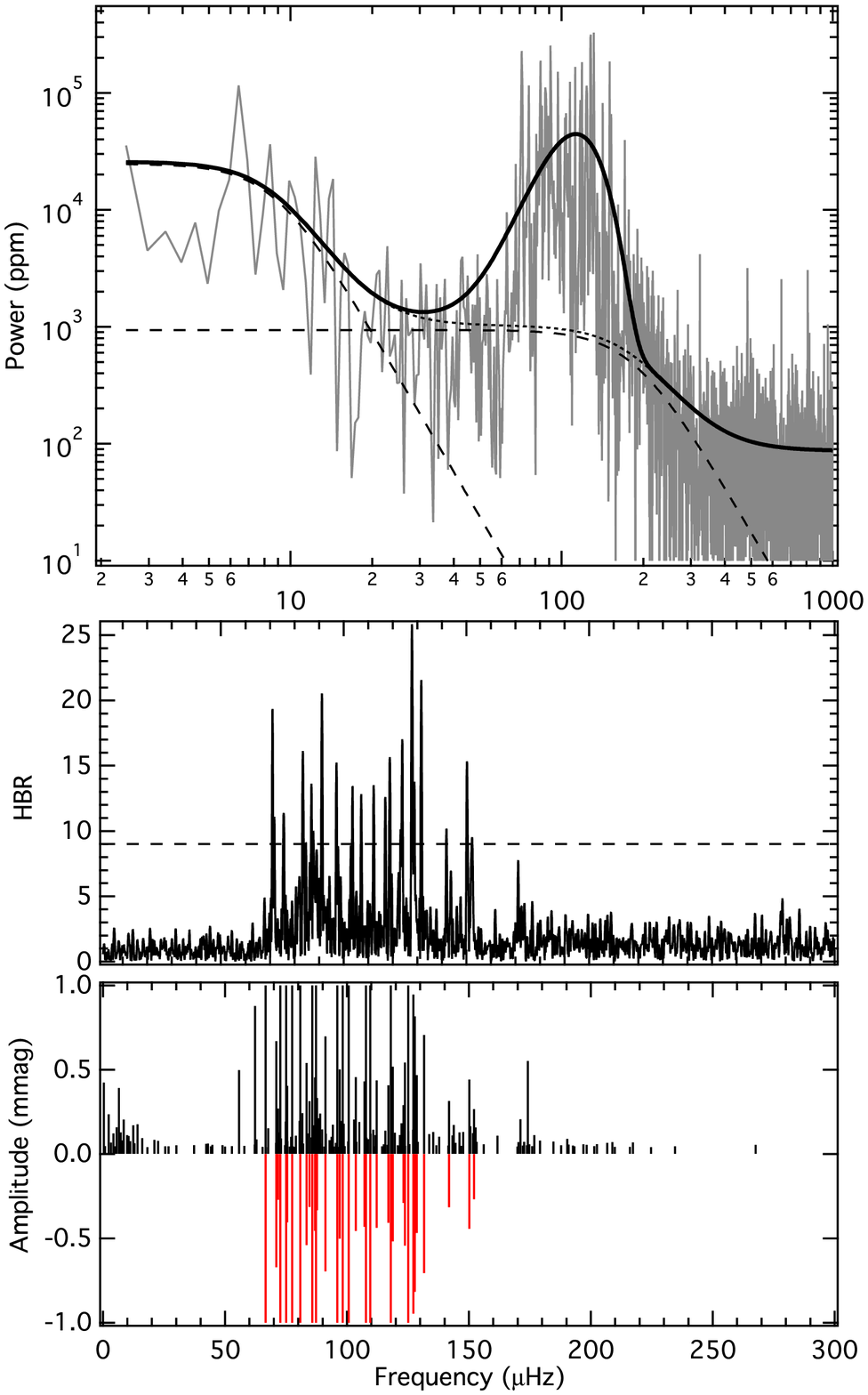}
\caption{{\it Top panel}: Residual gap-filled power spectrum of V 589 Mon (grey) after prewhitening the 15 highest amplitude frequencies. The global fit is shown as solid line, while the dotted line illustrates the models without the pulsational component consisting of the two Lorentzian-like functions (dashed lines) and white noise, and the dashed lines mark the two Lorentzian-like functions. {\it Middle panel}: Residual power spectra normalised to the background components, where the dashed line marks a height-to-background ratio (HBR) of 9 (given in power). {\it Bottom panel}: Comparison between the 197 formally significant frequencies (positive values) to the identified 37 pulsation frequencies (negative values) which consist of the 15 prewhitened peaks before granulation modeling and the 12 additional peaks exceeding an HBR value of 9.}
\label{v589gran}
\end{figure}

The same method as described for V 588 Mon was used for V 589 Mon. In this case the 15 frequencies with the highest amplitudes were prewhitened before the gaps in the residuals were filled by linear interpolation (Table \ref{v589-fs}). The top panel of Figure \ref{v589gran} shows the power spectrum of the gap filled data (grey), the global model fit (solid line), the model without the pulsational component (dotted line), and the two Lorentzian-like functions (dashed lines). 22 frequencies show an HBR value larger than 9 in power (middle panel in Figure \ref{v589gran}). Together with the 15 frequencies of highest amplitudes that were prewhitened, a total of 37 frequencies are attributed to be caused by pulsation (see Figure \ref{v589amps} and Table \ref{v589-fs}).

\subsection{Comparison of the MOST 2004 and 2006 data to the CoRoT data}
The MOST 2004 light curves of V 588 Mon and V 589 Mon have 59344 and 60386 data points, respectively. The noise levels in the original amplitude spectra computed from 0 to 100 \cd (i.e., 0 to 1157 $\mu$Hz) are 125 and 133 ppm (see Table \ref{stat}). 
In 2006 V 588 Mon and V 589 Mon were observed only less than half the time as for 2004, hence the light curves consist only of 23137 and 23113 data points, respectively, and the resulting noise levels from 0 to 100 \cd\, in the original amplitude spectra are 405 and 429 ppm (Table \ref{stat}). 

For each star a separate frequency analysis was conducted and the frequencies common to both data sets were then used for the further analysis and for comparison to the CoRoT data. The modeling of the background noise was not possible due to the lower quality of the MOST data compared to the CoRoT observations.

\begin{figure}[htb]
\centering
\includegraphics[width=0.7\textwidth]{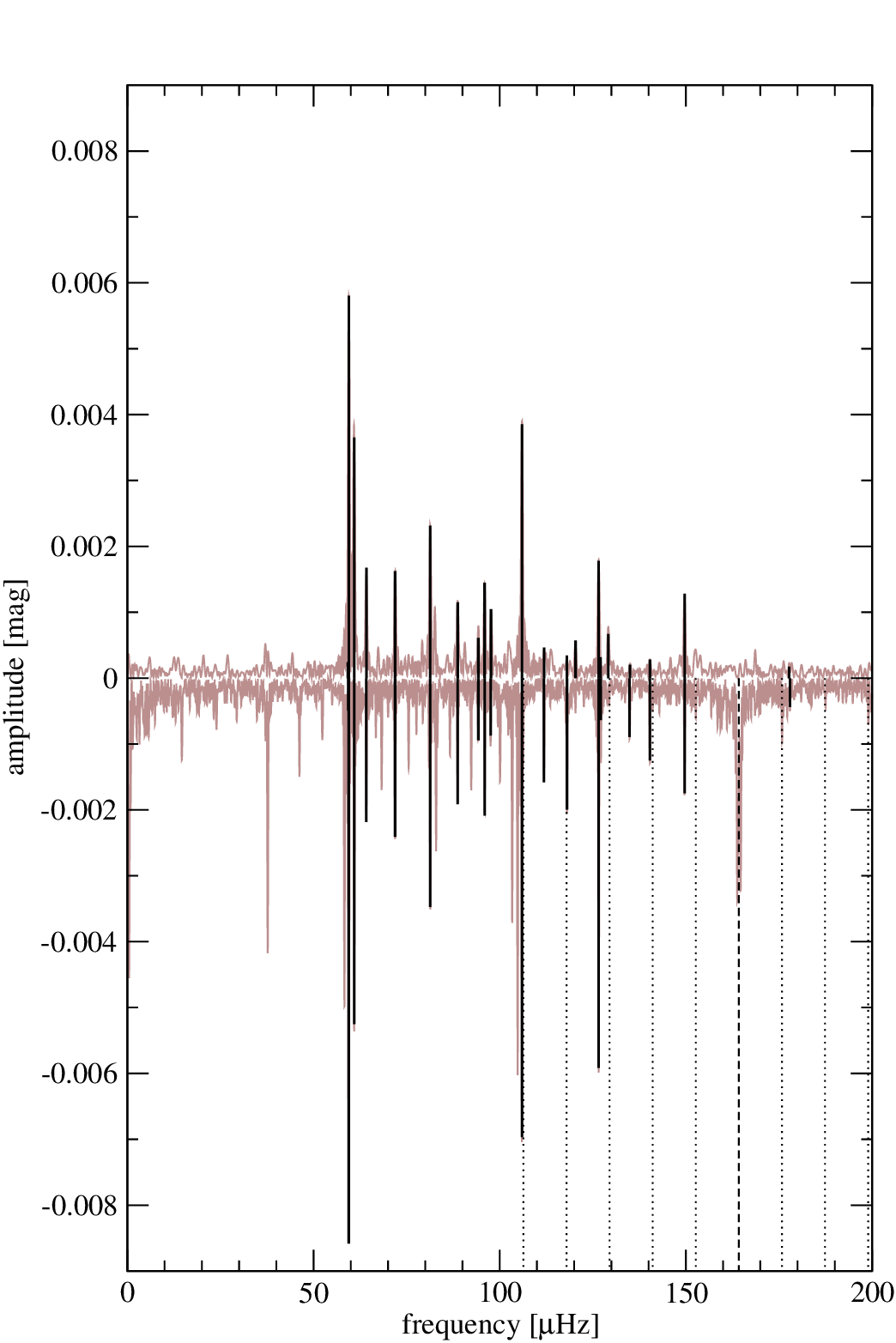}
\caption{Comparison of the V 588 Mon amplitude spectra (grey) derived from CoRoT (oriented upwards) and MOST 2004 (negative values assigned) data where the respective 21 and 19 pulsation frequencies are identified in black; the dashed line marks the MOST orbital frequency and the dotted lines are the respective 1 \cd sidelobes.}
\label{v588amps}
\end{figure}

For V 588 Mon 20 significant frequencies are common to the MOST data sets of 2004 and 2006. A cross identification with the 14 CoRoT frequencies that remain after granulation modeling showed that 12 of them are also present in the MOST data (Table \ref{v588-fs}). A further cross-check of the formally significant CoRoT frequencies {\it before} the background noise modeling to the 20 frequencies appearing in the MOST 2004 and 2006 data resulted in 7 additional common frequencies (F15 to F21 in Table \ref{v588-fs}). Although these peaks have HBR values lower than 9, hence would be attributed to be caused by background noise, they have to be considered as pulsational. Such stable frequencies that are present in three independent data sets obtained in three different years are unlikely to be caused by granulation. Additionally the shape of the pulsational power excess is very likely more complicated than a Gaussian that is used in the power spectrum model \citep{kal10} and in the future a more realistic approach will be needed. Therefore, for V 588 Mon 19 pulsation frequencies are in common between the MOST 2004, MOST 2006 and CoRoT data sets (Table \ref{v588-fs}) and are marked as black lines in the bottom panel of Figure \ref{v588amps}.
A comparison between the 106 formally significant frequencies and the identified 21 pulsation frequencies for V 588 Mon is given in the bottom panel of Figure \ref{v588gran} and illustrates the influence of granulation to the frequency spectrum.

\begin{figure}[htb]
\centering
\includegraphics[width=0.7\textwidth]{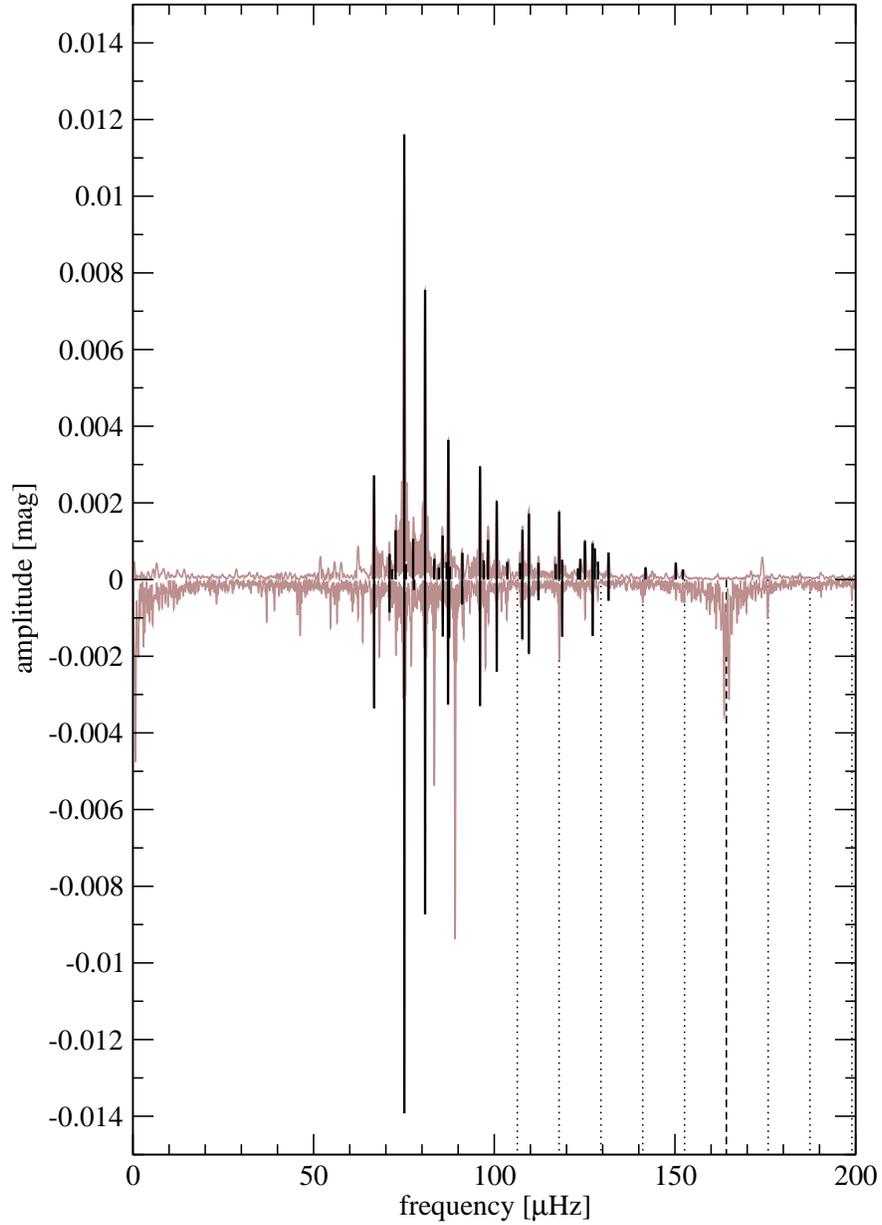}
\caption{Comparison of the V 589 Mon amplitude spectra (grey) derived from CoRoT (oriented upwards) and MOST 2004 (negative values assigned) data where the respective 37 and 17 pulsation frequencies are identified in black; the dashed line marks the MOST orbital frequency and the dotted lines are the respective 1 \cd sidelobes.}
\label{v589amps}
\end{figure}

For V 589 Mon 19 significant frequencies are common to the MOST data sets of both years. 17 of those were also identified in the CoRoT data after granulation modeling and are marked as black lines in the bottom panel of Figure \ref{v589amps}. The two additional frequencies appearing in the MOST data were identified likely to be aliases or caused by instrumental effects. The comparison between the formally significant 197 frequencies and the identified 37 pulsational peaks is shown in the bottom panel of Figure \ref{v589gran}.

{\it Using both MOST data sets and the COROT data, in total 21 pulsation frequencies are identified for V 588 Mon (black lines in the top panel of Figure \ref{v588amps}) and 37 for V 589 Mon (black lines in the top panel of Figure \ref{v589amps}) where 2 and 20 of them, respectively, are unique discoveries obtained with CoRoT. All MOST frequencies have a counterpart in the CoRoT observations.}

\section{Asteroseismic analysis of V 588 Mon and V 589 Mon}

For any seismic modeling of a star to be successful it is critical that the identified frequencies are, in fact, $p$-modes (or $g$-modes) intrinsic to the star. Errant or spurious frequencies are difficult to isolate by modeling alone. Because viable models are based on a large parameter space many possible solutions for various combinations and selections of the observed frequencies can be produced. In specific terms, when, as is often the case, it is impossible to find a model fit to all the observed frequencies, there usually exist multiple possibilities to fit some subset of the observed frequencies and no decisive way to prefer one solution over the others. 

CoRoT observations of both V 588 Mon and V 589 Mon have yielded a large number of frequencies, many of which were also seen by MOST \citep{gue09}. The two independent observations go a long way toward confirming the intrinsic, i.e., non-instrumental, nature of the frequencies. Furthermore, since the MOST and CoRoT observations are separated in time by several years, they imply the modes are stable over a period of at least 4 years.

Both stars show significant \vsini\, velocities, high enough, to produce resolvable splittings in the frequency domain. Unfortunately, the mode identification is complicated by the fact that the predicted splitting frequencies are comparable to the characteristic (i.e., asymptotic large) spacing that separates $p$-mode frequencies, thereby, making it difficult to distinguish the split sidelobes from the parent frequencies.
Furthermore, if the stars are rapidly rotating then the splittings are not linearly spaced about the reference frequency \citep{esp04}. Recent studies \citep{rees09b,deu10,oua08} show that non-perturbative multi-dimensional approaches are necessary to predict the frequency spacings. Although the direct modeling of these frequencies is complicated, \citet{rees09a} show how an empirical formula can be used to fit the frequencies assuming the interior rotation profile is simple enough. Needless to say, as long as mode identification depends on the interior rotation curve, there will always be some ambiguity in the actual identifications.

Finally, we note that the observed frequencies are low, corresponding to model modes near the fundamental frequency of the star. Although the frequencies of these modes are more sensitive to the interior structure of the star than higher frequency modes, hence, better suited to constraining the evolutionary state of the model, they are also subject to a mode bumping-like effect. As a consequence, the regular spacing between adjacent $p$-modes is perturbed, hindering mode identification.

In the next section we provide a general overview of the modeling methodology and theoretical considerations used to interpret the oscillation spectra. In sections 4.2. and 4.3. we discuss each star separately beginning with V 589 Mon.

\subsection{Method}

\subsubsection{Stellar mode and model calculation}
All of our modeling and seismic analyses are based on a dense set of models that cover the HR-diagram. The PMS models start above the birthline on the Hayashi track and extend to the ZAMS (zero-age main sequence). Model masses range from 0.81 \Msun\, to 4.99 \Msun\, in increments of 0.01 \Msun. The models themselves were constructed using the Yale Stellar Evolution Code \citep[YREC:][]{dem08} incorporating the latest opacities, nuclear physics, and equation of state. Each model is resolved into $\sim$2000 shells with two-thirds of the shells covering the outer envelope and Eddington gray atmosphere. We estimate the model frequency uncertainty to be ~0.1\%. The grid density is such that $\sim$2000 models lie within 2-sigma of each star's position in the HR-diagram, of the $\sim$400\,000 models computed in total \citep{gue09}. 

The adiabatic $p$-mode frequencies for $l$ = 0, 1, 2, and 3 of the models were computed using Guenther's stellar pulsation code \citep{gue04}. We have assumed that geometric cancelation will occur for higher order $p$-modes making them more difficult to see above the background noise level (see discussion in section 3.1).


\subsubsection{Mode bumping in PMS stars}
When the oscillation frequencies of our PMS models are plotted in an echelle diagram (i.e., frequency modulo the large spacing plotted opposite the frequency of the modes) they reveal the vertical alignment characteristic of common $l$-valued $p$-modes. But maybe surprising at first glance, the lowest radial order, $n$, modes do not stay aligned with the modes at higher frequencies but zigzag. This behavior is common to evolved post-main sequence stars and is attributed to the increased density gradients in the region surrounding the isothermal helium core near the base of the hydrogen burning shell. The effect is called mode bumping and is well documented in texts on stellar pulsation \citep[e.g.,][]{cox80, unn89}. PMS stars do not show the same large gradients in density because nuclear burning has not yet begun.

\begin{figure}[htb]
\centering
\includegraphics[width=8.5cm]{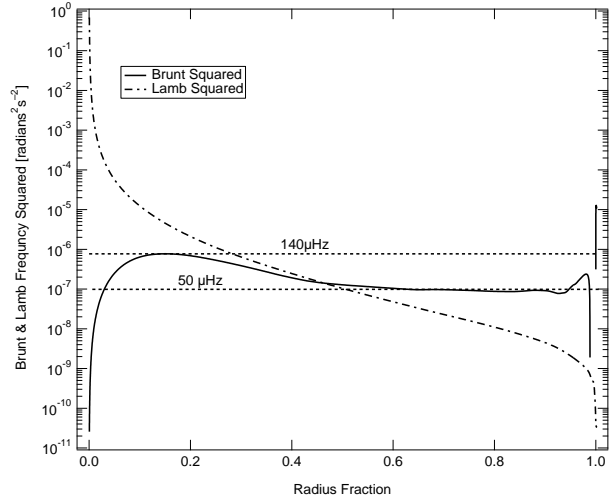}
\caption{Propagation diagram for the interior model of V 589 Mon. The square of the Brunt V\"ais\"al\"a and Lamb frequencies, in units of radians$^{\rm 2}$ s$^{\rm -2}$, are plotted against radius fraction.}
\label{DG04}
\end{figure}

\begin{figure}[htb]
\centering
\includegraphics[width=8.5cm]{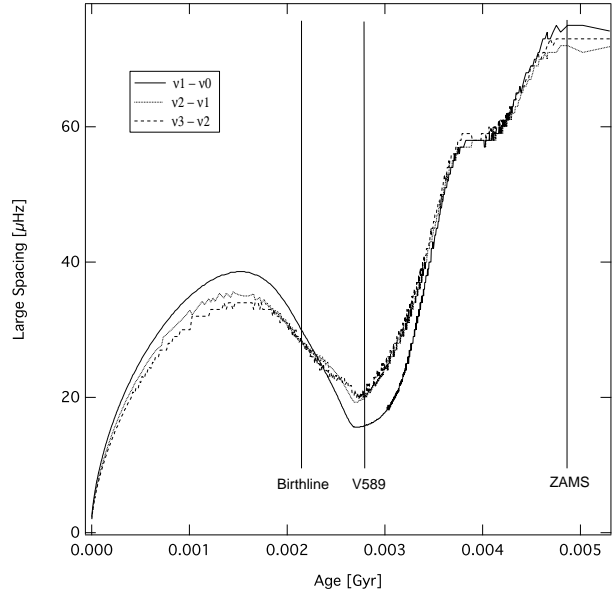}
\caption{The time evolution of the large frequency spacing of the lowest $n$-valued, $l = 0 p$-modes for a 2.65\Msun\, PMS model track (which passes through V 589 Mon's position in the HR-diagram). The frequency difference between the $n = 1$ and $n = 0$ $p$-modes is drawn with a solid line (indicated in the legend by $\nu1 - \nu0$). The frequency differences between the $n  = 2$ and $n = 1$, and $n = 3$ and $n= 2$ are drawn with dashed lines (as labeled in the legend). The birthline, the ZAMS, and the point where the track passes through V 589 Mon are indicated.}
\label{DG05}
\end{figure}

Regardless, for some PMS stars there exists a slight peak in the Brunt-V\"ais\"al\"a frequency in the deep interior (see Figure \ref{DG04} for V 589 Mon). This bump is enough to perturb the $p$-mode eigenfunction in the interior and affect its frequency. Note, only nonradial $p$-modes show mode bumping since they need to couple with $g$-modes (for which $l$ = 0 modes are undefined). The perturbation in frequency can be seen in a plot of $p$-mode frequency versus (evolutionary) time as shown in Figure \ref{DG05} for $l$ = 1 $p$-modes for models along a 2.65\Msun\,\,track (corresponding to V 589 Mon, see section 4.2.). The large spacing varies as the star evolves, changing inversely with the radius of the star. The $n$ = 0 large spacing deviates slightly from the large spacing of the other modes at various times during the evolution of the star.

Unlike the mode bumping that is seen in post-main sequence stars, the PMS bumping does not occur abruptly in time: it is only slowly varying as the model evolves toward the ZAMS. Consequently, it is not necessary to increase the resolution of the grid to follow this behavior as is the case of post-main sequence bumping. Although it complicates the identification of the modes at low frequencies, it also provides an additional feature of the oscillation spectrum that can be used to confirm the interior structure and evolutionary state of the PMS model.

\subsubsection{Searching for the best fitting model spectra}
To find the best-fit model oscillation spectrum to the observed spectrum we searched our grid of models \citep{guebro04} looking for local minima in $\chi^2$ defined by:
\begin{equation}
\chi^{2} = \frac{1}{N}{\sum_{i=1}^{N}} \frac{(\nu_{obs,i} - \nu_{mod,i})^2}{\sigma{^2}_{obs,i} + \sigma{^2}_{mod,i}}
\end{equation}
where $\nu_{obs,i}$ is the observed frequency for the $i^{\rm th}$ mode, $\nu_{mod,i}$ is the corresponding model frequency, $\sigma_{obs,i}$ is the observational uncertainty for the $i^{\rm th}$ mode, and $N$ is the total number of modes that match the observed frequencies. We estimate the model uncertainty $\sigma_{mod,i}$ by fitting models to the solar oscillation spectrum \citep{gue04}. Here we set it to 0.1\% the frequency of the mode.

\subsubsection{Rotation}
As we will describe in sections 4.2. and 4.3., we were unsuccessful in finding $p$-mode frequency fits to all the observed modes. Owing to the not insignificant rotation rates of the two stars, we considered the possibility that the oscillation spectrum of each star contains rotationally split modes. We included only the lowest order approximations in these attempts since full nonlinear rotational splitting computations are too laborious to be applied to a grid searching methodology. 

As a first attempt we considered a range of fixed width splittings. These are not model constructed splittings that depend on the eigenfunction of the mode and the interior structure but are simply a constant frequency added to and subtracted from the model frequency in order to simulate, approximately, the expected set of frequency splittings. In other words, when searching our grid of models we looked for frequency matches between model and observations at $\nu \pm m \Delta\nu$, where $\nu$ is a model computed $p$-mode, $m$ is the order of the splitting, and $\Delta\nu$ is the splitting frequency itself.

We also computed rotational splittings for a selected set of models, using the formulation of \citet{gou81}, for a solid body rotation curve and for a rotation curve that rises near the surface. Here we wanted to see if we could perceive any difference between the two different rotation curves, e.g., one curve produced slightly lower $\chi^2$ spectrum fits than the other. Note that Gough's formula is applicable to slow rotation rates only.

\subsection{V 589 Mon}
\subsubsection{Observational Constraints}
\begin{figure}[htb]
\centering
\includegraphics[width=8.5cm]{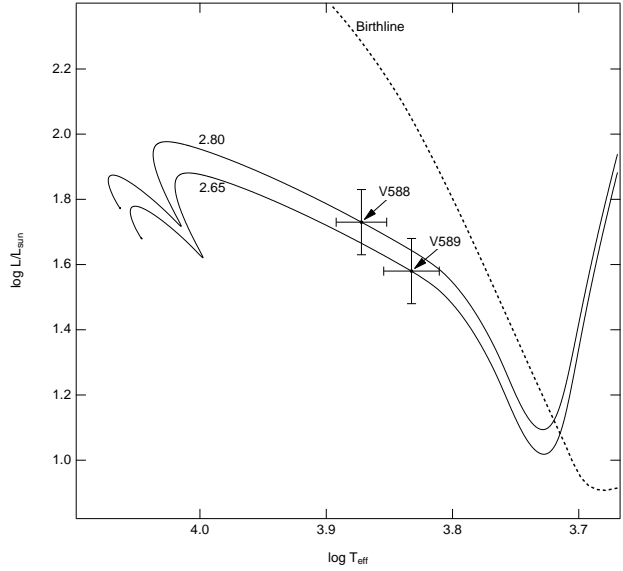}
\caption{Location of V 588 Mon and V 589 Mon in HR-diagram. The birthline, 2.65\Msun\, and 2.80\Msun\, pre-main sequence evolutionary tracks are shown.}
\label{DG01}
\end{figure}

V 589 Mon is located just below the birthline in the HR-diagram along our 2.65 \Msun\, PMS evolutionary track as shown in Figure \ref{DG01}. Note that its location in the HR-diagram and its association with the cluster NGC 2264 precludes it from being a post-main sequence star. The birthline \citep{sta83,pal99} corresponding to a mass in-fall rate of $\sim10^{\rm -5}$\Msun/year is shown. Following \citet{gue09} we take the luminosity of V 589 Mon to be log L/\Lsun\, = 1.58 $\pm$ 0.1, the effective temperature to be 6800 $\pm$ 350K, and we note that the observed \vsini\, is 60 $\pm$ 10 km s$^{\rm -1}$  \citep[see][their Table 2, values adopted from 2MASS photometry]{kal08}.

\subsubsection{Mode Analysis}
\begin{figure}[htb]
\centering
\includegraphics[width=8.5cm]{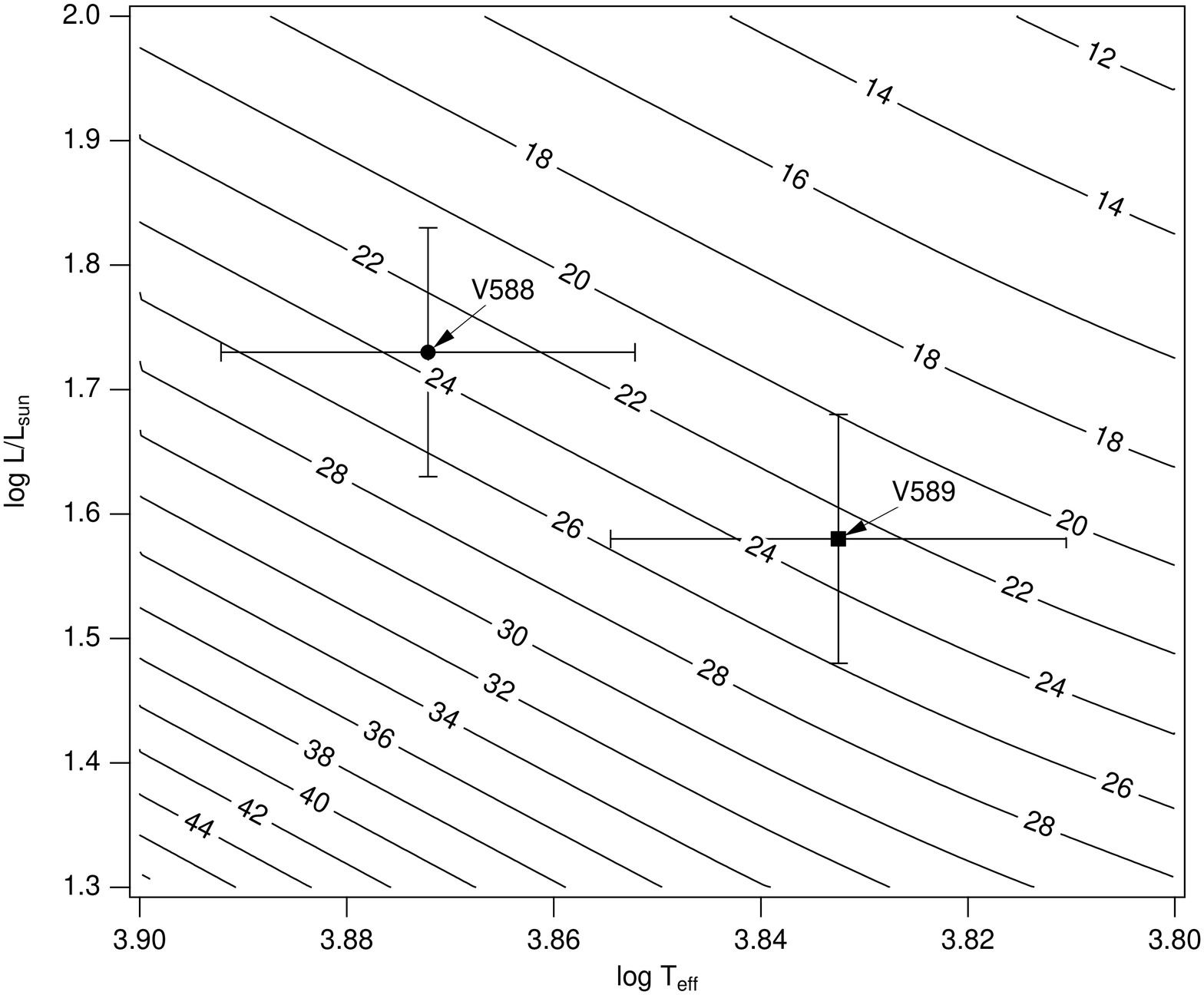}
\caption{Contour plot of characteristic spacing (asymptotic large spacing), in units of $\mu$Hz, computed from PMS models near V 588 Mon's and V 589 Mon's locations in the HR-diagram as indicated by the uncertainty bars. }
\label{DG02}
\end{figure}

Figure \ref{DG02} shows the model computed characteristic spacing \citep[approximately equal to the average large spacing in the asymptotic limit of large radial order $n$,][]{tas80} as function of position in the HR-diagram in the vicinity of  V 589 Mon (and V 588 Mon). The average small spacing, over the observed frequency range, computed from our model grid is 1.7 $\pm$ 0.3\,$ \mu$Hz.

We plot the observed frequencies in an echelle diagram to reveal the vertical alignment characteristic of common $l$-valued $p$-modes. Models show that the regular spacing between $p$-modes extends all the way to $n$ = 1 for PMS stars. Based on the model results shown in Figure \ref{DG02} we see that the large spacing for V 589 Mon should lie between 20 $\mu$Hz and 26 $\mu$Hz. The CoRoT frequencies have amplitudes down to 0.27 mmag where the MOST frequencies have amplitudes at or above 0.46 mmag. Note that this set of MOST frequencies include low amplitude frequencies that were excluded in the analysis of \citet{gue09} where only the 8 frequencies common to MOST and ground-based observations were presented and analyzed. We proceed with our model analysis using the CoRoT frequencies. 

\begin{figure}[htb]
\centering
\includegraphics[width=8.5cm]{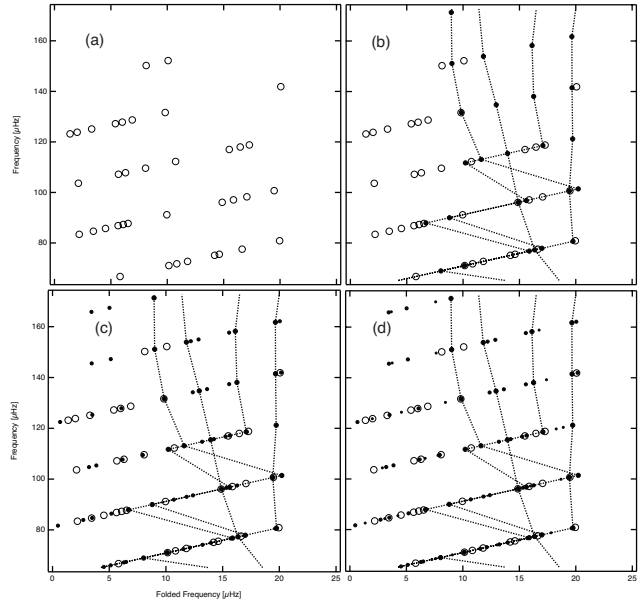}
\caption{Echelle diagrams for V 589 Mon with folding frequency 20.3$\mu$Hz. Each panel adds in sequence, starting with the CoRoT observed frequencies in (a), the model $l$ = 0, 1, 2, and 3 $p$-modes in (b), the $m = \pm 1$ model rotational split frequencies in (c), and the $m = \pm 2$ model rotational split frequencies in (d).}
\label{DG08}
\end{figure}

In an echelle diagram, there is some suggestion for vertically aligned sequences of modes in the observations, but the effect is confused by the presence of many additional apparently randomly scattered frequencies (see Figure \ref{DG08}). What strikes us most are the many more frequencies observed by CoRoT between 50 $\mu$Hz and 150 $\mu$Hz than can be accounted for from just $l$ = 0, 1, 2, and 3 $p$-modes. As noted in section 4.1 we believe geometric factors and the limited S/N of our data rule out the possibility of identifying higher $l$-values in our CoRoT (and MOST) data. 

We searched through all $\sim$400,000 PMS models in our grid comparing their oscillation spectra to the observed spectrum and identified those models that had $\chi^2 \leq 1$. We could find no model whose oscillation spectrum fit (with $\chi^2 \leq 1$) more than 9 of the 37 frequencies at a time with $l$ = 0, 1, 2, and 3 modes and which falls within the uncertainty box of V 589 Mon's position in the HR-diagram. The observed frequencies cannot all be accounted for by $l$ = 0, 1, 2, and 3 $p$-modes alone.

\subsubsection{Rotation}
\begin{figure}[htb]
\centering
\includegraphics[width=8.5cm]{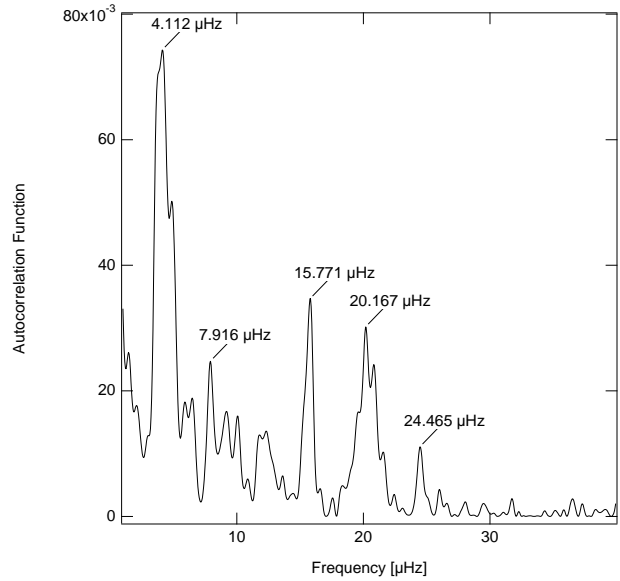}
\caption{Autocorrelation plot of the oscillation frequency spectrum of V 589 Mon observed by CoRoT with the frequencies of the highest peaks labeled.}
\label{DG06}
\end{figure}

\begin{figure}[htb]
\centering
\includegraphics[width=8.5cm]{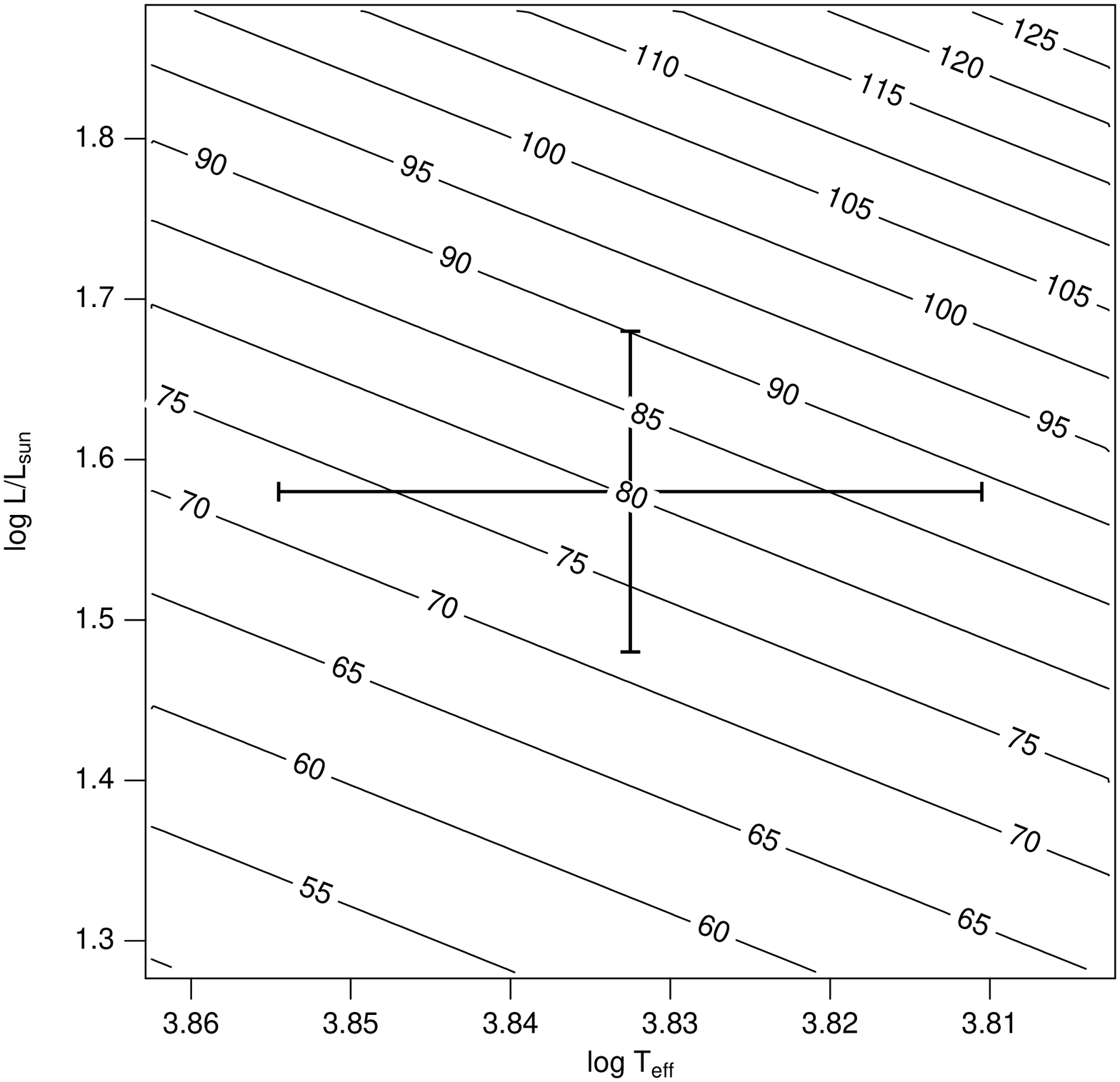}
\caption{Equatorial rotational velocities in units of km s$^{\rm Ð1}$ are shown in the vicinity of V 589 Mon's HR-diagram location. The rotational velocities are computed (they depend on the model radius) for a rotational splitting (solid body) of 4.1 $\mu$Hz.}
\label{DG07}
\end{figure}

To account for the extra modes present in the observed frequency range, we consider the possibility that rotationally split frequencies are present. An autocorrelation plot of the observed spectrum (Figure \ref{DG06}) shows a strong broad peak at 4.1 $\mu$Hz (f1), with lesser peaks at: 7.9 $\mu$Hz ($\sim$2f1), 20.2 $\mu$Hz (f2 $\sim$ large spacing), 15.8 $\mu$Hz (f2-f1), and 24.5 $\mu$Hz (f2+f1). We speculate that the 4.1 $\mu$Hz peak is due to $m = \pm 1$ rotational splittings and that the 7.9 $\mu$Hz peak is due to the $m = \pm 2$ rotational splittings. Recall that the small spacing, as determined from our models is $\sim$1.7 $\mu$Hz, hence, cannot explain the 4.1 $\mu$Hz peak, unless, of course, the deep interior structure of our models is grossly wrong. That the small spacing peak is not visible also supports our belief  that high $l$-valued modes are not present in this data set. The equatorial velocities, $v_{rot}$, corresponding to a 4.1 $\mu$Hz splitting produced by solid body rotation for models near V 589 Mon's location in the HR-diagram are shown in Figure \ref{DG07}. The rotation rate implied by the splitting frequency is consistent with the observed \vsini\, of 60 $\pm$ 10 km s$^{\rm -1}$. 

First to show how rotationally split modes fill up the echelle diagram, we computed rotational splittings, assuming solid body rotation (corresponding to $\sim 4 \mu$Hz splitting), for the model located closest to V 589 Mon's HR-diagram position, using the formulation from \citet{gou81}. In Figure \ref{DG08}  we show the CoRoT observations, (a) -- (d), along with the unsplit model modes, (b) -- (d), the $m = \pm 1$ splittings, (c) and (d), and the $m = \pm 1, \pm 2$ splittings. The number density and distribution of model modes are a best match when the rotational splittings are included. But, clearly, there is no direct correspondence between the model frequencies and the observations. 

We repeated our $\chi^2$ search but used oscillation spectra that include a fixed width rotational splitting, which we varied from 2.5 $\mu$Hz to 7.5 $\mu$Hz in steps of 0.1 $\mu$Hz. Restricting our search to $m = 0, \pm 1$ splittings, we could find, in the best case, matches to 22 of the 37 frequencies at one time and lie within the uncertainty box of V 589 Mon's position in the HR-diagram. The splitting frequency of the best case match is 4.5 $\mu$Hz. For $m = \pm 1, \pm 2$ splittings, we could find model fits to 24 of the frequencies again with a splitting frequency equal to 4.5 $\mu$Hz. If we assume that we are seeing rotational split modes, the small difference between fitting just $m = \pm 1$ and $m = \pm 1, \pm 2$ modes suggests that, as expected, $m = \pm 2$ (and $m = \pm 3$) rotationally split frequencies are not as prominent in the observations as the $m = \pm 1$ modes. 

\begin{figure}[htb]
\centering
\includegraphics[width=8.5cm]{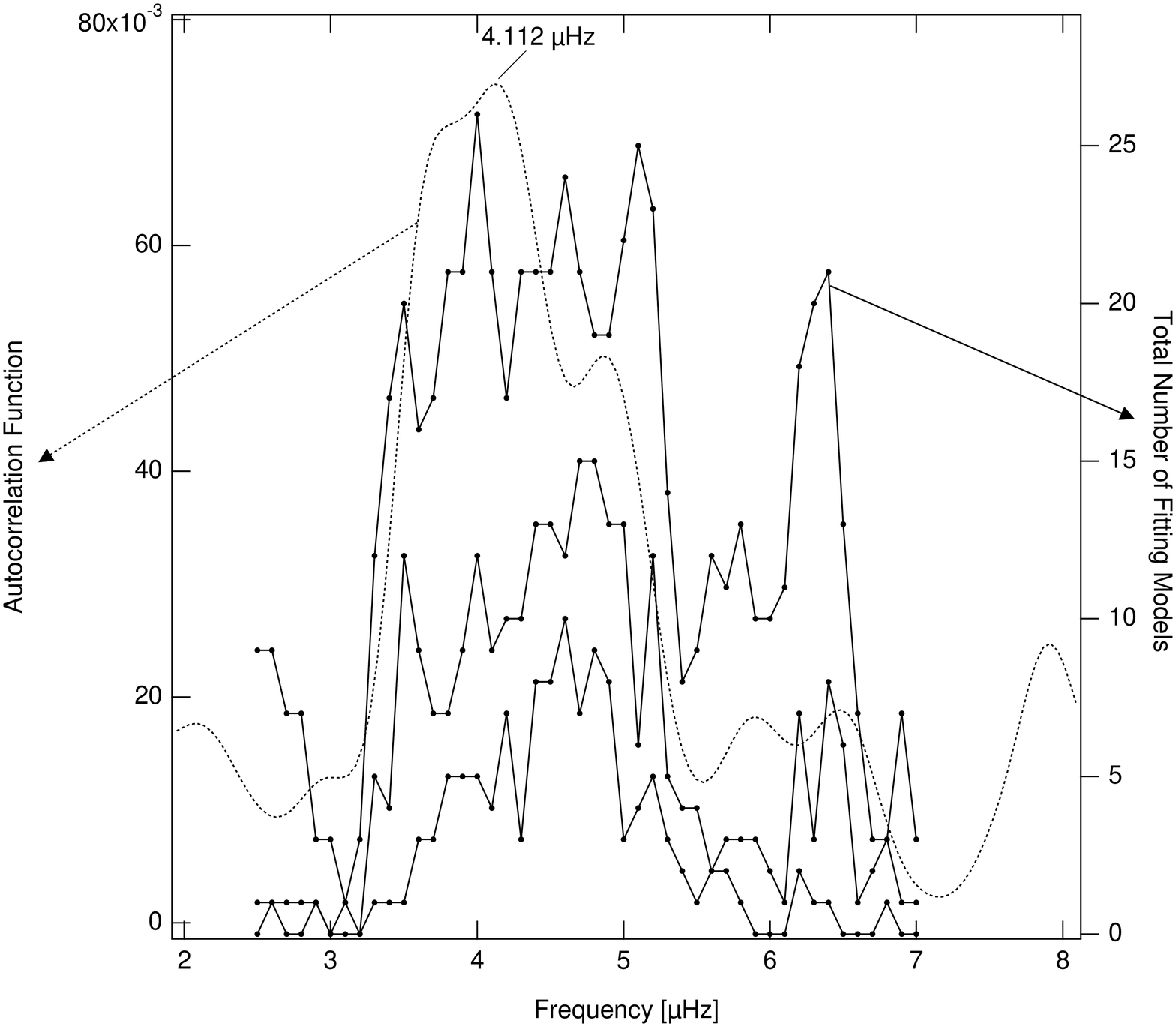}
\caption{The autocorrelation plot for CoRoT observations of V 589 Mon (dashed line) compared to the number of models that fit 17, 18, and 19 of the 37 observed frequencies for a given autocorrelation frequency ($\mu$Hz), and that lie within the uncertainty box of V 589 Mon's position in the HR-diagram. The lowest curve (solid line joining points) corresponds to the number of models fitting 19 frequencies. The next lowest curve corresponds to the number of models fitting 18 frequencies and the highest curve to the number of models fitting 17 frequencies.}
\label{DG09}
\end{figure}

In a crude attempt to show that the model fits do indeed prefer a splitting near 4.5 $\mu$Hz we computed the number of models whose oscillation spectra match exactly 17, 18, and 19 modes, of the 37 observed, and that also lie inside V 589 Mon's uncertainty box in the HR-diagram. When plotted as a function of the splitting frequency, Figure \ref{DG09}, we see that the number of models that fit the observations peaks for rotational splittings between 4 and 5 $\mu$Hz. The frequency dependence of the number of model fits matches the autocorrelation function. There is also a peak at 6.5 $\mu$Hz. This peak is likely the combined effect of the $l$ = 0 small spacing and the rotational splitting. Although we cannot extract any statistical quantities from this plot since the Model Count scale is arbitrary and depends on the density of models in our grid, it does show that models that have rotational splittings outside the range 4 $\mu$Hz to 5 $\mu$Hz do not fit the data as well as those within this range of splittings consistent with our interpretation that the observed spectrum contains sidelobes of rotationally split modes. 

That we are not able to fit all of the observed modes with our models may, in part, be due to the fact that we assumed the splitting frequency is constant, when, in fact, we know that the splitting does depend on model structure, the interior rotation curve, and the eigenfunction of the mode itself. To get a sense of how important these effects are we computed rotational splittings for a solid body rotation curve and for a rotation curve in which the angular velocity is constant throughout the interior then linearly increases by a factor almost a factor of two over the outer 2\% radius of the star, using the formula by \citet{gou81}, which is applicable to slow rotation rates.

\begin{figure}[htb]
\centering
\includegraphics[width=8.5cm]{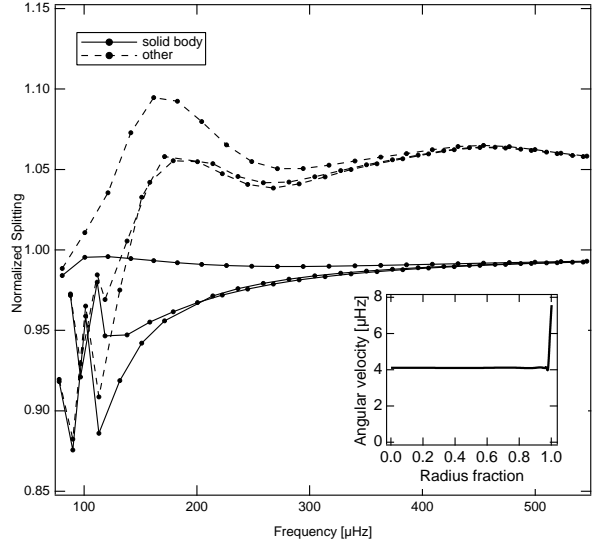}
\caption{Normalized rotational splittings ($\Delta\nu / \nu$) as a function of frequency in $\mu$Hz, $\nu$, computed from a model of V 589 Mon for solid body rotation (solid line) and for the rotation curve shown in the insert (dashed line).}
\label{DG10}
\end{figure}

In Figure \ref{DG10} we plot the splitting frequency normalized to the surface rotation rate frequency as a function of frequency for $l$ = 1, 2, and 3 $p$-modes.  The rotational splitting is constant except at the lowest frequencies where from mode to mode it varies by about 10\%. Coincidentally, this is the frequency range over which the oscillation frequencies for V 589 Mon are observed. At low frequencies the variation for a 4 $\mu$Hz splitting is $\sim$0.4 $\mu$Hz. Since we have taken our modeling uncertainty to be 0.1\%, which corresponds to frequency uncertainty of 0.1 $\mu$Hz for a 100 $\mu$Hz mode, our model fits should be sensitive to the variation in splitting with frequency. 

To test if we could see any difference in our $\chi^2$ searches, we computed a new model pulsation grid that includes the frequency-dependent splittings for each mode assuming a solid body internal rotation curve. For a solid body rotation rate corresponding to a 4.1 $\mu$Hz rotation rate at the equator of the star we did not find any significantly better or poorer model fits than our constant frequency splittings test at 4.1 $\mu$Hz. Due to the more laborious nature of these more refined model fits and their dependence on the internal rotation curve itself, we have not yet pursued more thorough searches over a wide range of rotation rates and rotation curves. We believe a different strategy is needed.

\subsection{V 588 Mon}
\subsubsection{Observational Constraints}
V 588 Mon is located just below the birthline in the HR-diagram along a 2.80 \Msun PMS evolutionary track, as shown in Figure \ref{DG01}. Following \citet{gue09} we take the luminosity of V 588 Mon to be log L/\Lsun = 1.73 $\pm$ 0.1, the effective temperature to be T$_{\rm eff}$ = 7450 $\pm$ 350 K, and the observed \vsini\, is 130 $\pm$ 20 km s$^{\rm -1}$  \citep[see][their Table 2, values adopted from 2MASS photometry]{kal08}. Except for the higher \vsini\, the star appears to be similar to V 589 Mon. 

\subsubsection{Mode Analysis}
Figure \ref{DG02} shows the expected large spacing as obtained from our model grid in the vicinity of V 588 Mon's location in the HR-diagram. Accordingly, we use a folding frequency of 22 $\mu$Hz in our echelle diagrams for V 588 Mon.  

\begin{figure}[htb]
\centering
\includegraphics[width=8.5cm]{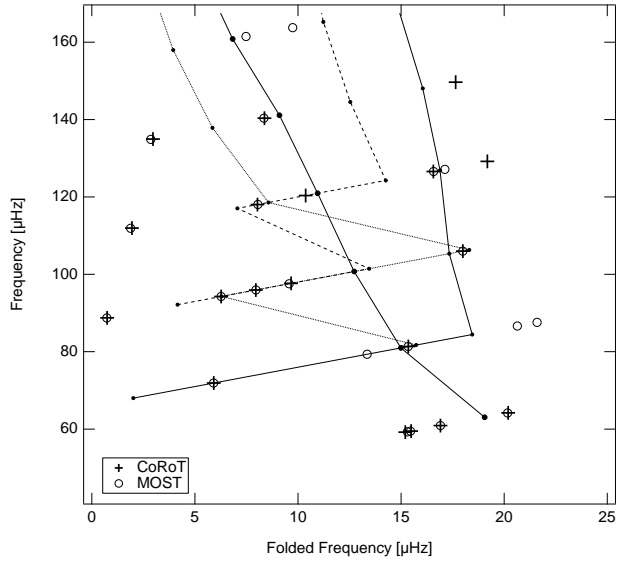}
\caption{Echelle diagram showing all the significant frequencies observed by CoRoT and MOST of V 588 Mon falling between 50 $\mu$Hz and 175 $\mu$Hz. The folding frequency is 22$\mu$Hz. The $l$ = 0, 1, 2, and 3 $p$-modes of the model that falls closest to V 588 Mon's position in the HR-diagram are indicated by a solid line connecting large points, a solid line connecting smaller points, a short dashed line connecting smaller points, and a long dashed line connecting smaller points, respectively.}
\label{DG12rev1}
\end{figure}

We identify 14 significant frequencies in the CoRoT observations above an amplitude of 0.5 mmag. MOST detects 24 significant frequencies above this amplitude. As with the MOST data for V 589 Mon, we have included all significant modes not just those selected by \citet{gue09} found in both ground and MOST data. The CoRoT and MOST sets of frequencies are shown in the echelle diagram of Figure \ref{DG12rev1}. Our model analysis is restricted to the CoRoT frequencies. We also plot the $l$ = 0, 1, 2, and 3 $p$-modes for the model in our grid that lies closest to V 588 Mon's location in the HR-diagram. 

There is some suggestion of vertical alignment in the observed frequencies as seen in the echelle diagram. But unlike V 589 Mon the distribution and density of modes does not point to an obvious over-abundance of frequencies. 

As with V 589 Mon we first sought model fits with $l$ = 0, 1, 2, and 3 $p$-modes to the observed frequencies (excluding rotational splittings). We were only able to find models lying with the HR-diagram uncertainty box for V 588 Mon that fit more than 4 (of the 14 CoRoT modes) with $\chi^2 \leq 1$. Indeed, for fits to 5 of the frequencies, no models within 2 sigma of the star's HR-diagram location were found. Again we are lead to believe the extra modes are rotational splittings.

\subsubsection{Rotation}
\begin{figure}[htb]
\centering
\includegraphics[width=8.5cm]{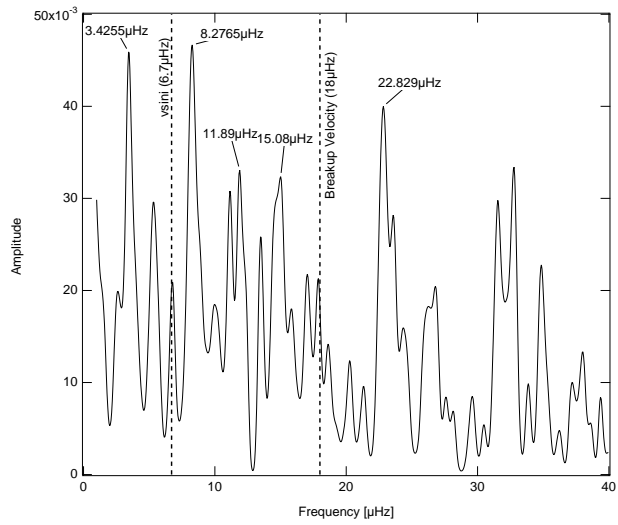}
\caption{Autocorrelation plot of the oscillation frequency spectrum of V 588 Mon observed by CoRoT with the frequencies of the highest peaks labeled. The lower limit to the splitting frequency set by the observed \vsini\,\, and the splitting frequency corresponding to the equatorial breakup velocity are also indicated.}
\label{DG13}
\end{figure}

Figure \ref{DG13} shows the autocorrelation function for V 588 Mon's oscillation spectrum. The observed \vsini\, sets a lower limit on the possible rotational splittings of $\sim$6.7 $\mu$Hz (assuming that V 588 Mon's location in the HR-diagram is correct) and the breakup velocity (where the centripetal acceleration at the equator equals the gravitational acceleration at the surface) sets an upper limit on the splitting of $\sim$18 $\mu$Hz (again assuming the model parameters based on V 588 Mon's location in the HR-diagram are correct). The autocorrelation plot does not reveal any obvious splitting peak, although the large spacing frequency does appear to be present at $\sim$23 $\mu$Hz. The possible splitting peaks are located at 8.3 $\mu$Hz, 11.9 $\mu$Hz, and 15.1 $\mu$Hz. But clearly none stand out. 

We carried out $\chi^2$ searches with fixed rotational splittings ranging from 5 $\mu$Hz to 18 $\mu$Hz but unlike V 589 Mon did not find any preferred frequency splitting or range of splittings. 
We believe that the rotation rate is, in fact, high enough for nonlinear effects to be important. At high rotation rates the centrifugal and Coriolis force terms become important and introduce non-axisymmetric splittings. \citet{deu10} have computed the oscillation spectra of rapidly rotating, massive (10\Msun) ZAMS stars using a full 2.5D evolution code \citep[see also][who first show the complications of rotational splittings for rapid rotation]{esp04}. Deupree and Beslin's results are enlightening as they show that the autocorrelation function (or, equivalently, as used in their case, a Fourier transform of the oscillation spectrum) fails to show rotational splitting frequencies (see their Figure 6) already for stars with rotation rates at 10\% breakup velocity. The rotational velocity of V 588 Mon as inferred from its \vsini\, is minimally at 30\% breakup velocity. If we assume their results apply to our 2.8\Msun\, PMS star then we should expect to see an autocorrelation plot without any predominant splitting peaks (as we do). \citet{deu10} also note that once the rotation rate is above 20\% breakup velocity the splittings from different modes begin to overlap each other further complicating mode identification. And so, our failure to find a model oscillation spectrum, even when including fixed value rotational splittings, may simply be a consequence of the far more complex nature of the splittings than our models are capable of addressing. 

Rather than attempting to fit the frequencies with the asymptotic-like formula of \citet{rees09a}, which they obtained from non-perturbative modeling of frequency splittings in rapidly rotating modes for simple rotation curves we are pursuing a different approach that does not depend on any prior model or rotation curve assumptions. We will report on our results in a future publication.

\section{Conclusions}
The two pulsating PMS stars V 588 Mon and V 589 Mon were observed by MOST in 2004 and 2006 for in total $\sim$70 days and in 2008 by the CoRoT satellite for 23.4 days during the Short Run SRa01. Detailed frequency analyses were conducted for all available data sets and compared to each other.

Our analysis illustrates that the frequency dependent intrinsic background signal, i.e. granulation, can explain a large number of significant peaks detected in PMS stars. Granulation modeling was conducted using a first order model where two Lorentzian-like functions, white noise and a Gaussian are combined. The resulting number of frequencies is more consistent with the expected number of low-degree $p$-modes observed in integrated light.
But it was also shown that the shape of the pulsational power excess is very likely to be more complicated than a Gaussian and that a more sophisticated approach would be needed in the future. This effect is illustrated by the fact that for V 588 Mon 7 frequencies at low amplitudes would have been suppressed by the background noise model, but appear significantly in all three satellite data sets, i.e. in data from MOST 2004, MOST 2006 and CoRoT. If frequencies are stable over a period of 4 years, they are unlikely to be caused by granulation.

After a comparison of the independent analyses of the MOST 2004 and MOST 2006 data sets to the CoRoT data, for V 588 Mon and V 589 Mon 21 and 37 frequencies, respectively, can be attributed to pulsation, among those the 8 and 12 previously published frequencies \citep{gue09}.

Even after granulation filtering, both V 588 Mon and V 589 Mon have more frequencies observed than can be accounted for by $l$ = 1, 2, and 3 $p$-modes. We investigated the possibility that the extra frequencies are rotational split modes. For V 589 Mon the autocorrelation of the observed frequency spectrum and our model searches support the notion that rotational split frequencies, with splittings between 4 $\mu$Hz and 5 $\mu$Hz, are present. We are able to match more than half of the observed frequencies with model fits. 
For V 588 Mon the autocorrelation plot and our attempts to fit model spectra that included splittings were unsuccessful. We believe this is because V 588 Mon's rotation rate is high enough (as implied by its \vsini) that its splittings are in the nonlinear regime as described by \citet{esp04}, and further studied by \citet{oua08}, \citet{rees09b}, and \citet{deu10}.

How can we obtain unambiguous mode identifications for V 588 Mon and V 589 Mon when we do not know the interior rotation curve? Is inversion possible, i.e., determining the rotation curve from the splittings, simultaneously with identifying the modes? Unlike the slowly rotating Sun, where the splittings are in the linear regime, V 588 Mon and V 589 Mon are rotating rapidly enough that non-perturbative models are probably necessary, especially for V 588 Mon. Rather than to continue with more sophisticated models we are currently pursuing a different approach in which we apply a Bayesean approach that incorporates model independent prior knowledge about the split and unsplit modes to estimate model fit likelihoods. We will report on our efforts in a future paper.


\acknowledgments
KZ acknowledges support by the Austrian {\it Fonds zur F\"orderung der wissenschaftlichen Forschung} (project T335-N16). KZ is recipient of an APART fellowship of the Austrian Academy of Sciences at the Institute of Astronomy of the University Vienna.
The Natural Sciences and Engineering Research Council of Canada supports the research of D.B.G.and M.G.; 
R.K. and W.W.W. are supported by the Austrian Research Promotion Agency (FFG).

\begin{deluxetable}{lrrccccc}
\tablewidth{0pt}
\tablecaption{Properties of the V 588 Mon and V 589 Mon data sets obtained from MOST in 2004 and 2006 and from COROT in 2008: time base (T), frequency resolution (fres), number of data points (N), point-to-point scatter (pt2pt), standard deviation (sigma) and noise level computed from the original data sets from 0 to 100 \cd (noise level).\label{stat}}
\tablehead{
\colhead{data set} & \colhead{star} &\colhead{T} & \colhead{fres} & \colhead{N} & \colhead{pt2pt} & \colhead{sigma} & \colhead{noise level} \\
\colhead{ } & \colhead{ } &\colhead{[d]} & \colhead{[\cd]} & \colhead{\#} & \colhead{[mag]} &
\colhead{[mag]} & \colhead{[ppm]} 
}
\startdata
MOST 04 & V 588 Mon & 48.13 & 0.02 & 59344 & 0.008 & 0.015 & 125 \\
                  & V 589 Mon & 48.13 & 0.02 & 60386 & 0.009 & 0.018 & 133 \\
MOST 06 & V 588 Mon & 22.77 & 0.04 & 23137 & 0.013 & 0.022 & 405 \\
                  & V 589 Mon & 22.77 & 0.04 & 23113 & 0.013 & 0.024 & 429 \\
CoRoT   & V 588 Mon & 23.41 & 0.04 & 56978 & 0.003 & 0.007 & 47 \\
                  & V 589 Mon & 23.41 & 0.04 & 57092 & 0.0009 & 0.010 & 61 \\
\enddata
\end{deluxetable}

\begin{deluxetable}{rrrcccr}
\tablewidth{0pt}
\tabletypesize{\small}
\tablecaption{V 588 Mon: 21 pulsation frequencies (freq) identified from the CoRoT data with their respective last digit errors given in parentheses, amplitudes (amp), height-to-background values (HBR) where appropriate, flag for frequencies also present in the MOST 2004 and 2006 data sets (MOST), and the cross-identification to the previously published 8 frequencies \citep[Gue09;][]{gue09}. The first 10 frequencies with the respective highest amplitudes were prewhitened before granulation modelling. \label{v588-fs}}
\tablehead{
\colhead{No.} & \colhead{freq} & \colhead{freq} & \colhead{amp} & \colhead{HBR} & 
\colhead{MOST}  & \colhead{Gue09}  \\
\colhead{\#} & \colhead{[\cd]} & \colhead{[$\mu$Hz]} & \colhead{[mmag]} &
\colhead{power } & \colhead{y / n }   & \colhead{\#}
}
\startdata
F1	&	5.1394(7)	&	59.484(8)	&	5.804	&		& y & F1	\\
F2	&	9.1579(9)	&	105.99(1)	&	3.853	&		& y & -	\\
F3	&	5.2628(8)	&	60.91(1)	&	3.653	&		& y & F3	\\
F4	&	7.029(1)	&	81.35(1)	&	2.315	&		& y & F4	\\
F5	&	10.935(1)	&	126.56(2)	&	1.779	&		& y & F2	\\
F6	&	5.546(1)	&	64.19(2)	&	1.673	&		& y & F5	\\
F7	&	6.213(1)	&	71.91(2)	&	1.624	&		& y & -	\\
F8	&	8.290(2)	&	95.95(2)	&	1.446	&		& y & F6	\\
F9	&	12.930(2)	&	149.65(2)	&	1.279	&		& y & -	\\
F10	&	7.668(2)	&	88.74(2)	&	1.152	&		& y & F7	\\
\tableline
F11	&	8.438(2)	&	97.67(2)	&	1.046	&	17.6	& y  & -	\\
F12	&	11.161(3)	&	129.18(4)	&	0.670	&	12.6	& n  & -	\\
F13	&	8.145(3)	&	94.27(4)	&	0.610	&	11.0	& y  & -	\\
F14	&	10.400(4)	&	120.37(4)	&	0.571	&	10.6	& n  & -	\\
\tableline
F15  &    9.672(4)     &      111.95(5)     &     0.464          &     7.6 & y  & -   \\
F16  &  10.198(6)    &      118.03(7)     &     0.344           &    6.1 & y   & -  \\
F17	&	10.981(6)	&	127.10(7)	&	0.317	&	6.7	& y & F8	\\
F18  &  12.128(7)    &      140.37(8)     &     0.280          &     5.1 & y   & -   \\
F19  &   5.115(8)      &     59.20(9)     &        0.243          &     4.3 & y  & -    \\
F20 &  11.661(9)     &     134.9(1)     &        0.203         &      4.1 & y  & -     \\
F21 &  15.36(1)       &     177.7(1)    &        0.175          &      4.4 & y   & -    \\
\enddata
\end{deluxetable}

\begin{deluxetable}{rrrcccr}
\tablewidth{0pt}
\tabletypesize{\scriptsize}
\tablecaption{V 589 Mon: 37 pulsation frequencies (freq) identified from the CoRoT data with their respective last digit errors given in parentheses, amplitudes (amp), height-to-background values (HBR) where appropriate, a flag for frequencies also present in the MOST 2004 and 2006 data sets (MOST) and the cross-identification to the previously published 12 frequencies \citep[Gue09;][]{gue09} The first 15 frequencies with the respective highest amplitudes were prewhitened before granulation modelling. 
\label{v589-fs}}
\tablehead{
\colhead{No.} & \colhead{freq} & \colhead{freq} & \colhead{amp} & \colhead{HBR} &
\colhead{MOST} & \colhead{Gue09}  \\
\colhead{\#} & \colhead{[\cd]} & \colhead{[$\mu$Hz]} & \colhead{[mmag]} &
\colhead{power} & \colhead{y / n }  & \colhead{\#}
}
\startdata
F1	&	6.4884(5)	&	75.097(6)	&	11.615	&		& y & F1	\\
F2	&	6.9862(6)	&	80.859(6)	&	7.551	&		& y & F2	\\
F3	&	7.5426(8)	&	87.299(9)	&	3.640	&		& n & - 	\\
F4	&	8.3005(9)	&	96.07(1)	&	2.954	&		& y & F3	\\
F5	&	5.7643(9)	&	66.72(1)	&	2.719	&		& y & F4 	\\
F6	&	8.698(1)	&	100.68(1)	&	2.048	&		& y & F5	\\
F7	&	10.192(1)	&	117.97(1)	&	1.773	&		& y & F6 	\\
F8	&	9.467(1)	&	109.57(1)	&	1.720	&		& y & F7 	\\
F9	&	9.314(1)	&	107.80(2)	&	1.299	&		& y & F9	\\
F10	&	6.281(1)	&	72.70(1)	&	1.287	&		& y & F8	\\
F11	&	7.407(1)	&	85.73(2)	&	1.147	&		& n & -	\\
F12	&	6.699(1)	&	77.54(2)	&	1.063	&		& y & F12 	\\
F13	&	8.490(1)	&	98.26(2)	&	1.036	&		& n & -	\\
F14	&	10.809(1)	&	125.10(2)	&	1.007	&		& n & -	\\
F15	&	10.990(1)	&	127.20(2)	&	0.945	&		& y & F11	\\
\tableline
F16	&	11.043(1)	&	127.81(2)	&	0.816	&	25.6	& n & -	\\
F17	&	11.372(2)	&	131.62(2)	&	0.706	&	21.6	& y & - 	\\
F18	&	7.876(2)	&	91.16(2)	&	0.696	&	20.5	& y & - 	\\
F19	&	6.138(2)	&	71.04(2)	&	0.670	&	19.3	& y & - 	\\
F20	&	10.697(2)	&	123.81(2)	&	0.543	&	17.0	& n & - 	\\
F21	&	7.205(2)	&	83.40(2)	&	0.540	&	16.1	& n & - 	\\
F22	&	10.262(2)	&	118.78(2)	&	0.517	&	15.7	& y & F10	\\
F23	&	8.386(2)	&	97.06(2)	&	0.501	&	15.2	& n & - 	\\
F24	&	11.119(2)	&	128.70(2)	&	0.468	&	13.8	& n & - 	\\
F25	&	8.953(2)	&	103.62(2)	&	0.456	&	13.4	& n & - 	\\
F26	&	7.503(2)	&	86.84(2)	&	0.455	&	13.6	& n & - 	\\
F27	&	12.980(2)	&	150.23(2)	&	0.442	&	15.3	& n & - 	\\
F28	&	9.698(2)	&	112.24(2)	&	0.437	&	13.5	& y & - 	\\
F29	&	9.260(2)	&	107.17(2)	&	0.431	&	12.8	& n & - 	\\
F30	&	10.109(2)	&	117.00(2)	&	0.408	&	12.6	& n & - 	\\
F31	&	6.524(2)	&	75.51(2)	&	0.405	&	11.4	& n & - 	\\
F32	&	7.579(2)	&	87.72(3)	&	0.333	&	10.0	& y & - 	\\
F33	&	12.257(2)	&	141.87(3)	&	0.314	&	10.2	& n & - 	\\
F34	&	7.313(3)	&	84.65(3)	&	0.313	&	9.1	& n & - 	\\
F35	&	10.643(3)	&	123.18(3)	&	0.290	&	10.1	& n & - 	\\
F36	&	6.199(3)	&	71.74(3)	&	0.270	&	11.1	& n & - 	\\
F37	&	13.148(3)	&	152.18(3)	&	0.267	&	9.5	& n & - 	\\
\enddata
\end{deluxetable}

\end{document}